# Enhanced Broadband Yellow Emission from Charge-Transfer States in TADF Organic Ionic Pair

Veerapandian Vallapandian[1], Rajaboopathi Mani[2], Sugasri Chinnasamy[1], and Alagiri Mani[1*]

[1]Sustainable Energy Materials and Interfaces Laboratory, Department of Physics and Nanotechnology, SRM Institute of Science and Technology, Kattankulathur-603203, Chengalpattu, Tamil Nadu, India.

[2]Key Laboratory of Emergent Materials, Department of Physics and Nanotechnology, SRM Institute of Science and Technology, Kattankulathur-603203, Chengalpattu, Tamil Nadu, India.

**Abstract**

Organic ionic pairs offer a promising route to thermally activated delayed fluorescence (TADF) by enabling small singlet–triplet energy gaps ($\Delta E_{ST}$) and tunable charge-transfer (CT) interactions. However, balancing between HOMO–LUMO decoupling and suppression of nonradiative decay remains a challenge. In this work, an electrostatically bound ionic pair, acridinium p-nitrobenzoate (AcPNB) exhibits a well-defined donor-acceptor (D−A) pair, and its physicochemical properties were studied. Single-crystal analysis reveals slipped $\pi$–$\pi$ stacking (3.4 Å) and a slightly twisted D−A geometry (dihedral angle of -11.86°), while strong N−H···O and C–H···O interactions stabilize CT states. Density functional theory (DFT) calculations reveal decoupled frontier molecular orbitals, with a small $\Delta E_{S1T1}$ of 6 meV and a SOC of 2.4 $cm^{-1}$. The AcPNB exhibits atypical broad photoluminescence (PL) emission centred at 578 nm, attributed to charge-transfer emission. The prompt and delayed lifetimes in *ns* and *μs*, together with temperature-dependent PL and broad emission confirm a charge-transfer excited state and characteristic of TADF mechanism. The cyclic voltametric studies experimentally confirm the charge-transfer process, and this value matches with DFT calculated HOMO value. The melting transition occurs at 171 °C, indicating strong ionic interactions against thermal decomposition. These finding insights into the ionic pair-based TADF organic emitter for OLED applications by exploiting the advantages of decoupled HOMO−LUMO and low $\Delta E_{ST}$.

## 1. Introduction

Organic materials have attracted significant attention owing to their versatility in electronic and optoelectronic devices, including organic field-effect transistors (OFETs), organic light-emitting diodes (OLEDs), organic light-emitting electrochemical cells (LEECs), organic solar cells, chemical, and biological sensors.[1,2] Among these technologies, OLEDs are considered highly promising for next-generation display and lighting applications due to their high energy-conversion efficiency, mechanical flexibility, and long operational lifetimes compared to conventional inorganic lighting.[3] The first practical OLED device was proposed by Tang and VanSlyke, in which small organic molecules were employed as emissive layers.[4] Early OLED research primarily utilized anthracene and acridine-based molecules as emitters owing to their favorable luminescent properties.[5] In conventional fluorescent OLEDs, only 25% of excitons those with singlet spin multiplicity can contribute to light emission, while the remaining 75% of triplet excitons are lost through nonradiative decay. To overcome this limitation and achieve near-complete exciton utilization, phosphorescent OLEDs were developed using transition-metal complexes and rare-earth materials. These approaches exploit strong spin-orbit coupling (SOC) induced by heavy metals, enabling efficient harvesting of both singlet and triplet excitons and thus approaching 100% internal quantum efficiency. Despite their high efficiency, phosphorescent materials often suffer from high cost, limited color stability, and potential environmental concerns, motivating the search for metal-free exciton-harvesting strategies. The triplet-triplet annihilation (TTA) offers an alternative and metal-free pathway for converting nonradiative triplet excitons into radiative singlet excitons, thereby extending exciton harvesting beyond the 25% spin-statistical limit of purely fluorescent OLEDs. TTA requires the interaction between two triplet excitons and typically occurs at high triplet densities. These high triplet–triplet interactions can also lead to non-radiative decay.[6,7] In this context, thermally activated delayed fluorescence (TADF) materials with donor-acceptor (D−A) systems were improved by Chihaya Adachi and co-workers in 2012, representing a breakthrough in organic light-emitting design.[8] TADF materials enable efficient reverse intersystem crossing (RISC) from the triplet to the singlet excited state by minimizing the singlet-triplet energy gap ($\Delta E_{ST}$), thereby allowing triplet excitons to contribute to delayed fluorescence emission.[9,10] In general, efficient TADF emitters are characterized by spatially separated highest occupied molecular orbital (HOMO) and lowest unoccupied molecular orbital (LUMO) distributions which reduce $\Delta E_{ST}$ in D−A or D−A−D molecular systems. Additionally, these materials typically exhibit short prompt fluorescence lifetimes associated with singlet excitons and long delayed lifetimes originating from triplet excitons.[11,12] In our previous work, we have strategically designed spatially well-separated HOMO-LUMO levels with a low $\Delta E_{ST}$ of less than 25 meV.[13] Three ionic pairs with positional isomers were reported: salicylic donors with acridine acceptors (Ac5ClSA, AcSA, and Ac3ClSA). However, achieving high quantum efficiency is challenging due to strong spin-orbit coupling and high rates of non-radiative transitions.

Ionic-interacted luminophores can be categorized into three types that are dominant in OLED applications: (1) ion-π interactions, (2) coordination interactions, and (3) electrostatic interactions.[14] The noncovalent ion-π interactions occur between ionic species and π-conjugated chromophores. Interactions between halide ions and π-conjugated aromatic systems have been found to induce strong SOC, which can drive improved RISC and triplet-state population. The conformational rigidification suppresses non-radiative decay.[15] Coordination-interacted systems can be constructed by metal-ligand coordination of aromatic acids with various metal ions (e.g., Zn, K, Cd) that have long-lived triplet excited states.[16] The electrostatically interacting systems tend to form an exciplex through intermolecular D−A components, which can lead to efficient TADF. The Coulombic attractions between oppositely charged molecular moieties efficiently restrict intramolecular vibrations and rotations, thereby suppressing non-radiative decay pathways. The conjugated D−A systems offer tunable frontier orbitals and controllable $\Delta E_{ST}$ gaps for the required emission.[17] The CT-dominated D−A molecules, including ionic-pairs can stabilize higher-lying excited states with extended lifetime for intense wideband emission with high quantum yield.[18] Several ionic D−A pairs were reported including CAZ-o-$BF_3^-$, dPhCAZ-o-$BF_3^-$ paired with TRZ-o-$ImEt^+$, TRZ-o-$ImEt^+$, TRZ-o-$ImEt^+$, and Py-X1, Qui-X2 (here X1=Br, $BF_4$, $PF_6$, Cl, I, and X2=Br, $BF_4$), as well as $Ac^+$ with $5ClSA^-$, $SA^-$, $3ClSA^-$. These are developed for efficient TADF.[18,19]

In this work, we have investigated a new TADF ionic pair acridinium p-nitrobenzoate (AcPNB) for enhanced yellow emission compared with our previously reported ionic yellow emitter AcSA. Both acridine and p-nitrobenzoic acid are widely investigated for optical and photophysical properties in laser technologies.[20] Acridine holds high electronegative units with relatively less extended π-conjugation and high triplet energy levels, which are beneficial for the CT process. Recently, some aryl-acridine emitters such as DMAC-TRZ (X), CNPh-X, $CF_3$Ph-X, dPh-X, BuPh-X, and OMePh-X were reported for TADF-OLEDs, and also anthracene-based systems were reported for TADF-LEECs.[21,22] By selecting p-nitrobenzoic acid as a donor, we observed that the PL intensity and lifetime of the excited charge carriers are significantly enhanced compared to our previously reported yellow emitter acridinium salicylate (AcSA).[13] Furthermore, the TADF, electronic, and photophysical properties were evaluated theoretically and experimentally. The PL emissions were observed in solid-state forms: powder, film, and 5 wt%, 10 wt%, and 15 wt% PMMA-doped films.

## 2. Experimental and theoretical studies

The precursors acridine (Ac) and p-nitrobenzoic acid (PNB) were obtained from Sigma-Aldrich and used as received, without further purification. Acridinium p-nitrobenzoate (AcPNB) was crystallized by a slow evaporation solution growth method. The equimolar ratios (1:1) of two precursors were dissolved separately in 15 mL of ethanol, after which the solutions were mixed and stirred at 50 °C for 3 h. The resulting pale-yellow and clear solution was filtered through Whatman filter paper to remove insoluble impurities. The filtrate was then left for slow evaporation at room temperature under ambient conditions. After 25 days, diamond-shaped block crystals were obtained for further characterization (see Figure S1). To understand the geometrical properties, quantum-chemical calculations were performed using density functional theory (DFT) and time-dependent DFT (TD-DFT). Ground and excited-state geometries were optimized using the B3LYP functional with the 6-311G$^{++}$ basis set to investigate the electronic structure of AcPNB. For Hirshfeld (HF) analysis, the CIF was used as an input file with the TONTO package integrated into Crystal Explorer.[23]

## 3. Results and discussion

### 3.1. Structure analysis and DFT calculations

The molecular structure of acridinium p-nitrobenzoate (AcPNB) is reported as an ionic pair formed via proton transfer from the donor carboxylic acid group to the nitrogen in the acridine acceptor.[24] The crystal structure is stabilised by $\pi$–$\pi$ stacking interactions between acridinium moieties with a centroid-to-centroid separation distance of 3.4 Å (for AcSA – 3.6 Å). The anionic p-nitrobenzoate is positioned in a perpendicular orientation with respect to the cationic acridinium framework with a dihedral angle of -11.86° as shown in Figure 1(a) and (b). This interlocked arrangement facilitates effective electrostatic and intermolecular interactions contributing to charge redistribution within the ionic pair. Within the crystal lattice, the perpendicular p-nitrobenzoate (PNB) donor interacts with two neighboring acridinium moieties through the $COO^-$ group. The protonated carboxylate C−O interacts with the $N-H^+$ site of one acridinium unit, while the C=O group associates with the C−H bond positioned opposite the nitrogen as the two acridinium units are positioned antiparallel to each other. Additionally, the nitro group of PNB participates in weak intermolecular contacts with the C−H bond located adjacent to the ring to the $N-H^+$ site of acridinium. Furthermore, the C−O group of PNB aligns nearly parallel with the C−H bond with an interactive distance of 3.6 Å next to the nitro group of another PNB molecule (Figure S2). These tightly zigzag or ladder-shaped (Figure S2 (a) and (b)) packing structural arrangements via π−π and C−H···O interactions strengthen the overall packing and provide a rigid molecular structure to facilitate CT and prolong their lifetimes.[13] Furthermore, the HF 3D surfaces were constructed to analyze intermolecular interactions in the crystal structure, as shown in Figure S3. The distances from HF surfaces to the nearest nucleus inside the surface ($d_i$), nearest nucleus outside the surface ($d_e$), and normalized contact distance ($d_{norm}$) are shown in Figure S3(a-c).,The 2D fingerprint plots were presented in Figure S3(d-h). The $d_{norm}$ surface involves a color scale in which red, white, and blue regions correspond to intermolecular contacts shorter than, equal to, and longer than the sum of van der Waals radii, respectively. Here, the red regions indicate the presence of close intermolecular contacts, which can be attributed to interactions such as N−H···O and C−H···O. The blue regions represent relatively weaker interactions. The associated $d_i$ and $d_e$ surfaces further reveal that these contacts are primarily localized around the aromatic rings and oxygen-containing functional groups, suggesting the presence of π−π interactions and C−H···O contacts. These features indicate that intermolecular interactions are directional and concentrated within specific regions of the molecule, contributing to the overall packing and intermolecular coupling in the solid state. The associated 2D fingerprint plots further quantify these interactions, showing that H···H contacts contribute the largest fraction (30.4%), followed by C···H (26.5%) and O···H (26.0%) interactions, while N···H interactions are comparatively minor (2.4%). The significant H···H contribution reflects the presence of weak van der Waals interactions and structural flexibility, which may facilitate nonradiative decay through vibrational relaxation.[25,26]

DFT calculations were carried out to gain information about electronic and excited-state properties, and the energy values are tabulated in Table 1. The frontier molecular orbitals analysis reveals that the HOMO is primarily localized on the donor (p-nitrobenzoate) moiety with minor extension toward the adjacent C–H region of the acridinium unit, whereas the LUMO is distributed over the acceptor (acridinium) unit, mainly distributed over the carboxylate group. From the HOMO −4$^{th}$ level, most area of the nitro group is involved in full HOMO occupation. The HOMO−LUMO overlap integral was evaluated and found to be essentially zero, indicating negligible spatial overlap and a highly decoupled D−A electronic structure. Furthermore, the SOC value was calculated between the singlet and triplet states and tabulated in Table S1. The calculated singlet-triplet energy gap for $\Delta E_{S1-T1}$ is 6 meV, and the SOC is 2.41 $cm^{-1}$. The SOC analysis reveals moderate $S_1$–$T_1$ coupling, while higher-lying excited states exhibit relatively stronger SOC, indicating that RISC is likely mediated through higher excited states rather than a direct $T_1 \rightarrow S_1$ pathway. The larger SOC values are observed between higher-lying states, particularly $S_2$-$T_2$ (36.02 $cm^{-1}$) and $T_2$–$S_3$ (34.49 $cm^{-1}$). Therefore, the higher-lying states with high SOC can behave like a heavy atom effect for effortless transition from $T_1$-$S_1$ states, but most of the transition occurs from $S_1$-$T_1$. This combination of low $\Delta E_{ST}$ and reasonably high SOC facilitates an efficient RISC process and thus TADF.[27,28]

**Table 1. Computed Values of HOMO-LUMO Energies and their Differences**

| | HOMO (eV) | LUMO (eV) | $E_g$ (eV) | $E_S$ (eV) | $E_T$ (eV) | $\Delta E_{S1T1}$ (meV) |
|---|---|---|---|---|---|---|
| **AcPNB** | − 6.73 | − 3.57 | 3.16 | 2.580 | 2.574 | 6 |

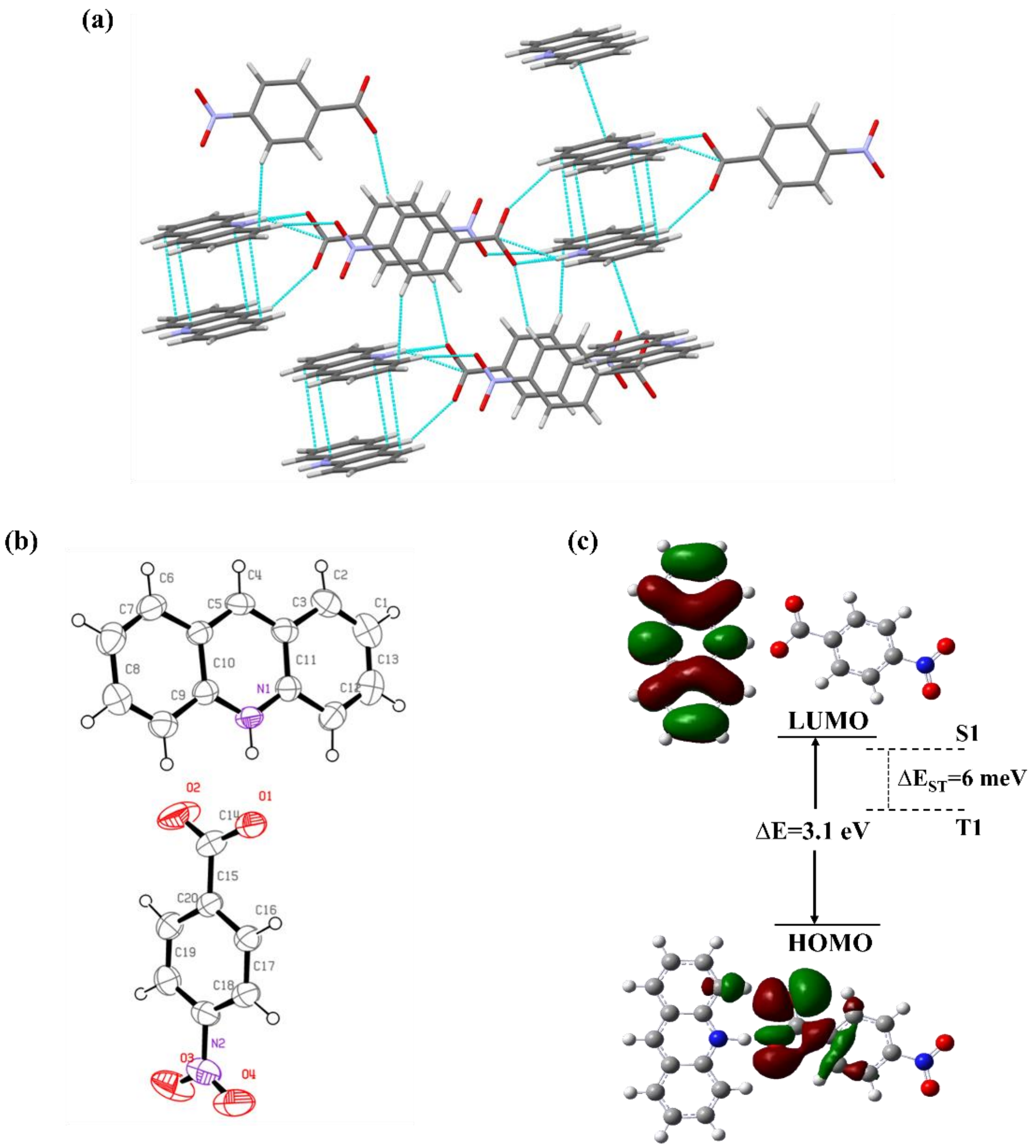

**Figure 1: (a)** π–π interactions, **(b)** ORTEP view, and **(c)** frontier molecular orbital distribution (down – HOMO and up – LUMO) of acridinium p-nitrobenzoate (AcPNB).

### 3.2. Photophysical and vibrational studies

The solid-state UV-Vis absorption spectra and the corresponding Tauc plot are presented in Figures 2(a) and 2(b), respectively. The weak absorption observed in the 200 to 235 nm region is attributed to π→π* electronic transitions, whereas the broad band spanning from 236 to 425 nm is primarily associated with n→π* transitions.[29,30] From the Tauc plot analysis, the optical band gap of AcPNB is estimated to be 2.5 eV consistent with its emission in the yellow region. The steady-state photoluminescence (PL) spectrum and corresponding CIE chromaticity diagram of the ionic pair are shown in Figures 2(c) and (d). Upon excitation at 355 nm, the AcPNB exhibits atypical broad-band and single emission centred at 578 nm. This suggests a strong intermolecular CT interaction between D−A, with no localized emission from exciton dissociation (LE).[18,31] The broad solid-state PL emission originates from the distribution of CT states energetically stabilized by π−π* and electrostatic environments within the ionic pair. A small $\Delta E_{ST}$ of 6 meV and sufficient SOC of 2.4 $cm^{-1}$ allow a spin-forbidden transition between $T_1 \rightarrow S_1$ states and thus RISC leading to the TADF-assisted broadband spectrum.[32] The PLQY is nearly 2% obtained. This low PLQY can be attributed to strong π−π stacking interactions, which promote intermolecular electronic coupling and facilitate exciton quenching. Similar behavior has been observed in previously reported ionic TADF systems such as Py-X1 and Qui-X2. Notably, Qui-based systems exhibit relatively strong π−π stacking with distances of 3.43 Å and 3.35 Å and relatively low PLQYs of 15% and 17% for Qui-Br and Qui-$BF_4$, respectively. Furthermore, another series of ionic systems, MeQ1-X and MeQ2-X (X = OTf, I, and $BF_4$), was reported. Among these systems, MeQ1-OTf exhibited a low PLQY (4%) due to its strong π−π intermolecular interactions and weak oscillator strength.[19,33] The small $\Delta E_{ST}$ and moderate SOC favour efficient RISC, strong intermolecular π–π and C−H···π interactions may promote broad emission, but their exciton delocalization and nonradiative decay limit the PLQY. Furthermore, our recently reported yellow-emitter AcSA exhibited dual emission due to excited-state intramolecular proton transfer (ESIPT) between hydroxyl (OH) and carboxylate ($COO^-$) groups, as well as the CT process (Figure S8). In contrast to AcSA, the AcPNB ionic system does not facilitate ESIPT; instead, the photophysical behaviour is governed by a CT process owing to the absence of a hydroxyl group in the p-nitrobenzoate donor. To further investigate the effect of the molecular environment on the photophysical properties, the ionic pair was examined in the film state in PMMA matrices with polymer concentrations of 5 wt%, 10 wt%, and 15 wt%, as shown in Figure S4. The polymer matrix was employed to regulate molecular rigidity and to stabilize electrostatic interactions between the anionic donor and the cationic acceptor. Furthermore, the improved PLQY was observed from 2% to 5.2% for the neat AcPNB thin film and 6.2% for the 10 wt % PMMA-doped film.

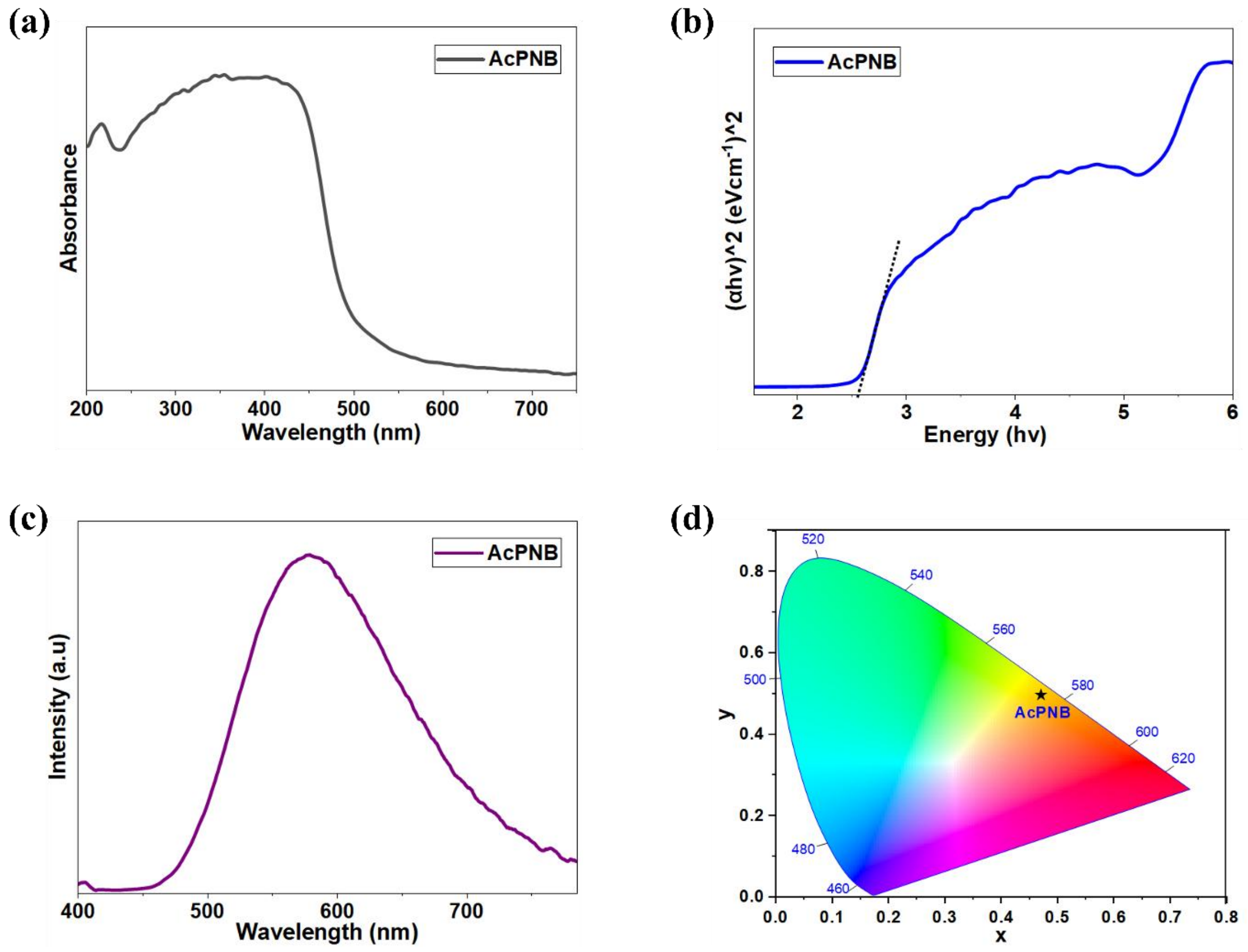

**Figure 2: (a)** Absorption spectra, **(b)** Tauc plot, **(c)** emission spectra, and **(d)** CIE chromaticity diagram of AcPNB (the coordinates are 0.47, 0.49).

The suppression of molecular bond vibrations upon incorporation into the PMMA matrix is corroborated by Raman spectroscopic analysis (Figure S5). As the polymer concentration increases, the molecular vibrations are progressively suppressed indicating a reduction in non-radiative decay pathways. Consequently, the optimized polymer-embedded AcPNB exhibits an enhanced PL intensity.[34,35] Raman vibrational analyses of pristine AcPNB and PMMA-doped thin films are presented in Figure S5. The vibrations of C−H, C−C−H, C−N, $NO_2$, and C=C functional groups are observed at 773, 1015, 1281, 1350, and 1487 $cm^{-1}$, respectively, for the pristine AcPNB. A prominent band centred at 1412 $cm^{-1}$ is attributed to the symmetric stretching vibration of the carboxylate group ($COO^-$). The observed vibrational frequencies are shifted relative to PNBA due to strong intermolecular interaction with acridinium, and the symmetric $COO^-$ vibration confirmed the proton transfer.[36,37] In the PMMA-doped films, additional peaks appear at 632, 861, and 1750 $cm^{-1}$ are originating from PMMA vibrations, while the peak at 1579 $cm^{-1}$ becomes broadened, which is attributed to the disordered polymer environment.[38] Notably, the intensities of these vibrational modes are significantly suppressed in the polymer-doped films indicating restricted molecular motions. These reduced vibrations play a critical role in promoting TADF by minimizing non-radiative decay pathways and facilitating efficient RISC.

Time-resolved photoluminescence measurements of solid-state AcPNB were carried out as shown in Figures 3 (a) and (b). The lifetime values and the quality fitting factor are listed in Table S2. The material exhibits a prompt fluorescence lifetime of 7.7 ns and a delayed lifetime of 10.2 μs, confirming TADF behavior with long-lived excited states in the ionic pair. As AcPNB has low $\Delta E_{ST}$, single emission and moderate SOC, it can be presumed that the delayed emission can be due to RISC from $T_1 \rightarrow S_1$ instead of $T_1 \rightarrow S_0$. The improved excited-state stabilization and radiative recombination efficiency of AcPNB is due to the p-nitrobenzoate donor that provides moderate HOMO overlap with the acceptor.[13] Compared to our previously reported yellow-emitting ionic pair AcSA, the AcPNB exhibits a higher photoluminescence intensity (see Fig S8) along with extended excited-state lifetimes. Specifically, AcPNB shows a prompt lifetime of 7.7 ns and a delayed lifetime of 10.2 μs, whereas AcSA displays shorter lifetimes of 4.0 ns and 2.8 μs, respectively (see Table S2).

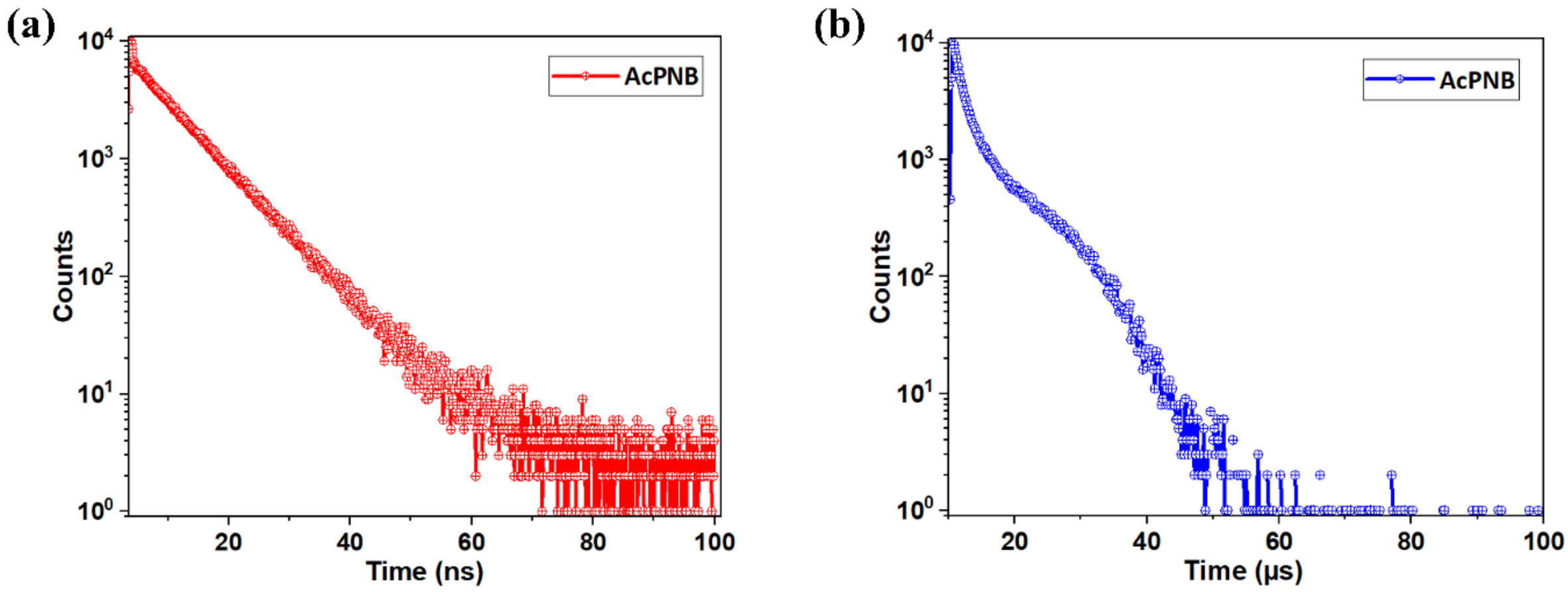

**Figure 3: (a)** Prompt lifetime and **(b)** delayed lifetime curves of AcPNB.

Furthermore, the temperature-dependent photoluminescence (PL) was investigated to further elucidate the TADF characteristics. The PL, temperature vs intensity plots were presented in Figure 4(a) and (b). The PL intensity decrease from 300 K to 250 K originates from suppressed reverse intersystem crossing (RISC) efficiency due to reduced thermal energy. Upon further cooling, the intensity increases at 200 K and subsequently declines at lower temperatures. This enhancement arises from the effective suppression of vibrationally induced non-radiative processes, thereby improving exciton confinement and radiative recombination.[39] At low temperatures, 150 K to 75 K, the PL intensity decreases again due to significant suppression of thermally activated RISC, leading to accumulation of triplet excitons that predominantly decay via non-radiative pathways. Overall, the observed non-monotonic variation in PL intensity reflects a delicate balance among internal conversion, radiative decay, CT-state stabilization, and RISC processes. Such behavior is consistent with temperature-dependent photophysical characteristics reported for TADF systems.[39–41]

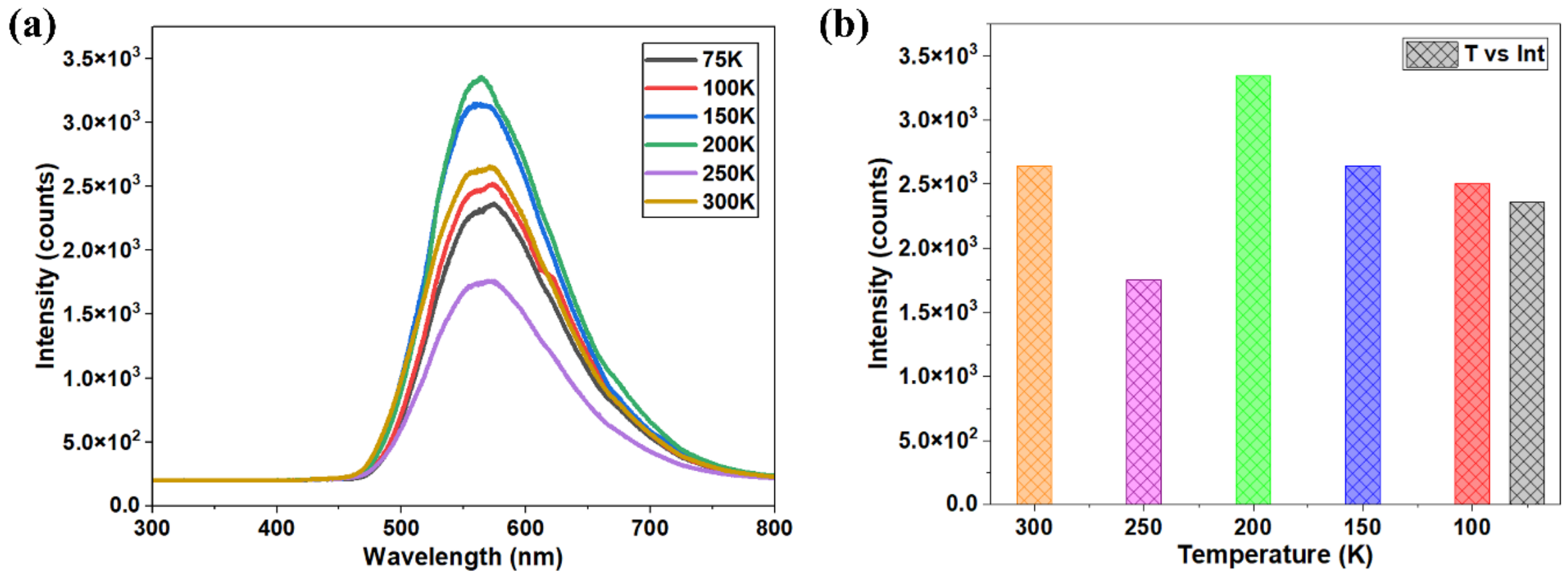

**Figure 4: (a)** Temperature-dependent photoluminescence (PL) spectra at different temperatures for acridinium p-nitrobenzoate (AcPNB) and **(b)** bar diagram of PL intensity at different temperatures.

The solvatochromic behavior of AcPNB was investigated in a series of solvents with varying polarities to elucidate intra- and intermolecular interactions as well as solute-solvent effects.[42] The selected solvents are based on moderately polar, polar protic, and polar aprotic categories, including dichloromethane (DCM), methanol (MeOH), ethanol (EtOH), acetonitrile (ACN), dimethylformamide (DMF), and dimethyl sulfoxide (DMSO). The UV and PL spectra of AcPNB recorded in different solvents are shown in Figure S6. In all solvents, a broad absorption band extending from 200 to 400 nm is observed, indicating comparable $\pi \rightarrow \pi^*$ and $n \rightarrow \pi^*$ electronic transitions regardless of solvent polarity. In contrast, the emission behavior is strongly solvent-dependent. In intermediate polar solvents such as ACN and DMF, multiple emission features are observed. Specifically, ACN shows a dominant emission at 472 nm with additional bands at 530 and 568 nm, while DMF exhibits a primary emission at 438 nm along with a broad emission spanning 520−610 nm. These emissions are attributed to radiative decay from contact ion pairs (CIP), associated with relatively localized excited states, and solvent-separated ion pairs (SSIP), which possess stronger CT character.[43] In contrast, a single broad emission band is observed in DCM, DMSO, EtOH, and MeOH, indicating the predominance of a single emissive excited state. Ion-pair equilibria can rationalize this solvent-dependent behavior: in intermediate-polarity solvents (ACN and DMF), partial solvation enables the coexistence of CIP and SSIP, resulting in multiple emission bands, whereas in highly polar or strongly solvating environments, dielectric screening stabilizes the SSIP state, leading to a single, red-shifted CT emission.[44,45] The AcPNB dissolved solvents and spin-coated films were examined under a UV chamber, and result given in Figure S7.

### 3.3. Electrochemical and thermal analysis

The cyclic voltammetry measurement for AcPNB is presented in Figure 5(a). The observed oxidation and redox features can evidence the CT processes between the HOMO and LUMO of the ionic pair. A well-defined oxidation peak appears at a potential of 0.27 V, indicating the loss of hydrogen, here the p-nitrobenzoate unit. The HOMO energy level estimated from electrochemical data shows good agreement with the value obtained from DFT calculations. For instance, the theoretically calculated HOMO level is −6.7 eV, whereas the experimentally derived value is −5.5 eV, demonstrating reasonable consistency between experimental and computational approaches. The deviation observed between HOMO values may be attributed to the stronger electron-withdrawing nature of the nitro ($NO_2$) group, which enhances ion-pair interactions and solvation effects, leading to greater stabilization of the oxidized states in solution. This correlation validates the reliability of the electrochemical analysis for evaluating the frontier molecular orbitals of AcPNB.[46–48]

The thermal stability of the AcPNB ionic pair was investigated by thermogravimetric analysis (TGA) and differential scanning calorimetry (DSC), as shown in Figure 5(b). The TGA profile shows no noticeable mass loss up to 135 °C indicating good thermal stability in this temperature region. A single-step mass loss is observed in the temperature range of 140 to 330 °C, indicating a dominant thermal decomposition associated with the gradual breakdown of functional moieties such as $COO^-$, $NO_2$, and $NH^+$ groups within the ionic pair. Beyond this temperature, e.g., 400 °C, the residual mass is attributed to the formation of a carbonaceous char arising from incomplete decomposition of the organic components. The DSC thermogram exhibits a sharp and narrow endothermic peak is attributed at 171 °C to the melting transition of the ionic pair at 171 °C. There is no glass transition observed before the melting point, indicating the crystallinity of the sample. At higher temperatures, a weak and broad thermal feature is observed between 296 and 336 °C, which may be attributed to thermally induced structural relaxation or minor molecular rearrangements. Beyond 350 °C, the material undergoes thermal decomposition, consistent with the observed mass loss in the TGA profile.[49,50]

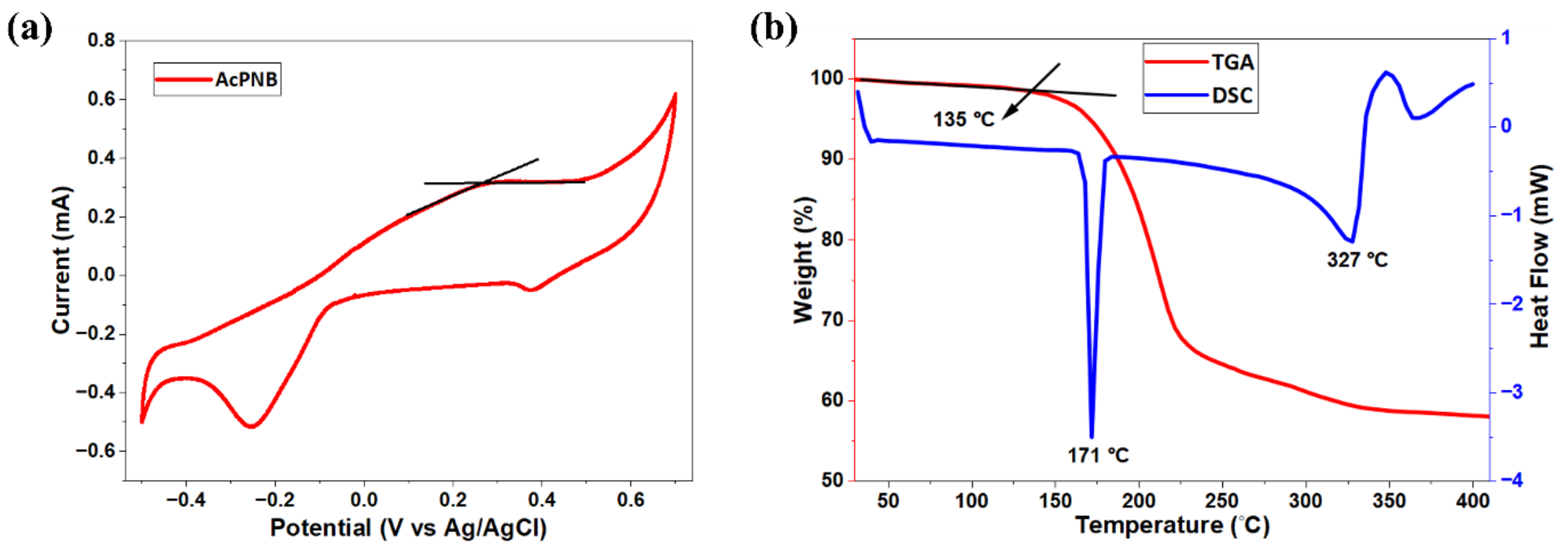

**Figure 5: (a)** Cyclic voltammogram at 10 mV/s and **(b)** TGA, DSC plot of AcPNB.

## 4. Conclusions

In summary, acridinium p-nitrobenzoate (AcPNB) was demonstrated as an efficient yellow-emitting ionic pair, outperforming the previously reported acridinium salicylate (AcSA). The AcPNB was successfully crystallized via a slow solvent evaporation method and systematically investigated for its optical, electrochemical, and thermal properties. The observation of a single, broad, and consistent PL emission in the powder, film, and PMMA-doped thin films confirms a dominant CT excited state stabilized by balanced electrostatic interactions within the ionic pair. The AcPNB exhibits PLQY ≈ 2% in the solid state, and PLQY improved to 5.2 % for AcPNB and 6.2 % for 10 wt % PMMA-doped thin films. DFT and TD-DFT calculations reveal spatially separated HOMO and LUMO, with a low $\Delta E_{ST}$ = 6 meV and a SOC of 2.4 cm-1, supporting the feasibility of TADF. This is further confirmed by time-resolved photoluminescence measurements, which exhibit a prompt lifetime of 7.7 ns and a delayed decay time of 10.2 µs. Furthermore, the observed temperature-dependent PL (300−75K) behaviour supports TADF, governed by the interplay of intersystem crossing and thermally assisted RISC alongside competing radiative and non-radiative decay pathways. The excited-state strengths, such as locally excited (LE) and charge-transfer, and solute-solvent interactions were studied by solvatochromism studies. Electrochemical analysis shows good agreement between experimentally determined HOMO energy levels, indicating effective intra- and intermolecular CT. Thermal analysis confirms adequate stability with a well-defined decomposition onset above 140 °C. Although the present study demonstrates that simple ionic pairs can achieve low $\Delta E_{ST}$ and moderate spin–orbit coupling, we are working on molecular design to suppress nonradiative decay and to enhance the internal quantum efficiency suitable for OLED applications.

**Author contributions**

Veerapandian Vallapandian: Writing original draft, editing, experimental, characterization, data curation, and conceptualization. Rajaboopathi Mani: Writing, editing, conceptualization, and formal analysis. Sugasri Chinnasamy: review, validation, and formal analysis. Alagiri Mani: Writing, editing, review, validation, project administration, conceptualization, data curation, and formal analysis.

**Conflicts of interest**

The authors declare no competing financial interests.

**Supporting information**

The data supporting this article have been included as part of the Supplementary information (SI): characterization techniques; SOC and time decay values; the digital photograph of the crystal and fluorescent thin-films; π−π interaction; Hirshfeld surfaces; Raman spectra for thin-films; PL emission of AcPNB in solid and liquid states.

**Acknowledgements**

The authors gratefully acknowledge the financial support and the High-Performance Computing Cluster (HPCC) for computational study from the Department of Physics and Nanotechnology, SRM IST. Additionally, acknowledge the Nanotechnology Research Centre (NRC) for providing characterization facilities and Pondicherry University for the Raman facility. The author R.M. acknowledges SRMIST for their funding support under the selective excellence research initiative (No: SRMIST/R/AR(A)/SERI2024/174/48). The authors sincerely acknowledge the Indian Nanoelectronics Users' Programme Idea to Innovation (INUP-i2i), IIT Bombay, for providing the temperature-dependent PL measurement facility.

**References:**

1 M. Rabiei, M. Hosseinnezhad, R. Ghahary, V. Nutalapati, S. Nasiri, J.-M. Nunzi, J. Padgurskas and R. Rukuiza, *ACS Applied Optical Materials*, 2025, **3**, 1483–1498.

2 Z. Qin, C. Gao, W. W. H. Wong, M. K. Riede, T. Wang, H. Dong, Y. Zhen and W. Hu, *J. Mater. Chem. C Mater.*, 2020, **8**, 14996–15008.

3 N. Sun, C. Jiang, Q. Li, D. Tan, S. Bi and J. Song, *Journal of Materials Science: Materials in Electronics 2020 31:23*, 2020, **31**, 20688–20729.

4 C. W. Tang and S. A. Vanslyke, *Appl. Phys. Lett.*, 1987, **51**, 913–915.

5 J. Huang, J. H. Su and H. Tian, *J. Mater. Chem.*, 2012, **22**, 10977–10989.

6 S. Nasiri, M. Rabiei, H. Shaki, M. Hosseinnezhad, K. Kalyani, A. Palevicius, A. Vilkauskas, G. Janusas, V. Nutalapati, S. Kment and J. Michel Nunzi, *J. Photochem. Photobiol. A Chem.*, 2024, **447**, 115289.

7 J. Wang, F. Meng, W. Liu, Z. Zhang and J. Li, *Mater. Chem. Front.*, 2024, **8**, 4106–4113.

8 H. Uoyama, K. Goushi, K. Shizu, H. Nomura and C. Adachi, *Nature 2012 492:7428*, 2012, **492**, 234–238.

9 X.-K. Chen, D. Kim and J.-L. Brédas, *Acc. Chem. Res.*, 2018, **51**, 2215–2224.

10 T. Zhang, Y. Xiao, H. Wang, S. Kong, R. Huang, V. Ka-Man Au, T. Yu and W. Huang, *Angew. Chem. Int. Ed.*, DOI:10.1002/anie.202301896.

11 T. Kalab, A. T. Gogoulis, B. O. Patrick and Z. M. Hudson, *Chemistry – A European Journal*, DOI:10.1002/chem.202502760.

12 M. K. Etherington, J. Gibson, H. F. Higginbotham, T. J. Penfold and A. P. Monkman, *Nat. Commun.*, 2016, **7**, 13680.

13 V. Vallapandian, R. Mani, I. B. Rietveld, A. Muhammed Ashraf and A. Mani, *ACS Applied Optical Materials*, 2025, **3**, 2603–2615.

14 W. Dai, Y. Jiang, Y. Lei, X. Huang, P. Sun, J. Shi, B. Tong, D. Yan, Z. Cai and Y. Dong, *Chem. Sci.*, 2024, **15**, 4222–4237.

15 G. Jiang, J. Yu, J. Wang and B. Z. Tang, *Aggregate*, DOI:10.1002/agt2.285.

16 X. Yang and D. Yan, *Adv. Opt. Mater.*, 2016, **4**, 897–905.

17 S. Chakraborty, R. Arikkottira M and R. K. Vijayaraghavan, *ACS Applied Optical Materials*, 2025, **3**, 2775–2791.

18 L. He, R. Bai, R. Yu, X. Meng, M. Tian and X. Wang, *Angew. Chem. Int. Ed.*, 2021, **60**, 6013–6020.

19 Z. Xu, D. Hean, C. Climent, D. Casanova and M. O. Wolf, *Mater. Adv.*, 2021, **2**, 5777–5784.

20 M. Pordel, H. Gheibi and A. Sharif, *Journal of Fluorescence 2024 35:7*, 2024, **35**, 4997–5034.

21 E. Crovini, K. Stavrou, P. Sahay, B. M. Nguyễn, T. Comerford, S. Warriner, W. Brütting, A. Monkman and E. Zysman-Colman, *The Journal of Physical Chemistry C*, 2024, **128**, 14429–14441.

22 M. Rabiei, R. Ghahary, M. Hosseinnezhad, A. Palevicius, A. Vilkauskas, G. Janusas, S. Nasiri, J. M. Nunzi, J. Padgurskas and R. Rukuiza, *Org. Electron.*, 2025, **141**, 107214.

23 P. R. Spackman, M. J. Turner, J. J. McKinnon, S. K. Wolff, D. J. Grimwood, D. Jayatilaka and M. A. Spackman, *J. Appl. Crystallogr.*, 2021, **54**, 1006–1011.

24 H. Balasubramanian, P. R. Mariappan and K. Poomani, *Acta Crystallogr. C Struct. Chem.*, 2024, **80**, 115–122.

25 R. Mani, I. B. Rietveld, B. Nicolaï, K. Varadharajan, M. Louhi-Kultanen and S. Narasimhan, *Chem. Phys.*, 2015, **458**, 52–61.

26 M. Ashfaq, K. S. Munawar, M. N. Tahir, N. Dege, M. Yaman, S. Muhammad, S. S. Alarfaji, H. Kargar and M. U. Arshad, *ACS Omega*, 2021, **6**, 22357–22366.

27 Y. He and D. Escudero, *Chem. Sci.*, DOI:10.1039/D6SC01726F.

28 A. Morgenstern, J. Weiser, L. Schreier, K. Gabel, T. Gabler, A. Ehm, D. Beer, N. Schwierz, U. T. Schwarz, K. Zeitler, C. Deibel, D. R. T. Zahn, C. Wiebeler and G. Salvan, *Adv. Opt. Mater.*, 2026, **14**, e01504.

29 R. Mani, I. B. Rietveld, K. Varadharajan, M. Louhi-Kultanen and S. Muthu, *Journal of Physical Chemistry A*, 2014, **118**, 6883–6892.

30 R. Bhowal, S. Biswas, A. Thumbarathil, A. L. Koner and D. Chopra, *The Journal of Physical Chemistry C*, 2018, **123**, 9311–9322.

31 W. Li, Y. Pan, L. Yao, H. Liu, S. Zhang, C. Wang, F. Shen, P. Lu, B. Yang and Y. Ma, *Adv. Opt. Mater.*, 2014, **2**, 892–901.

32 X. Lv, Y. Wang, N. Li, X. Cao, G. Xie, H. Huang, C. Zhong, L. Wang and C. Yang, *Chemical Engineering Journal*, 2020, **402**, 126173.

33 A. Klimash, A. Prlj, D. S. Yufit, A. Mallick, B. F. E. Curchod, P. R. McGonigal, P. J. Skabara and M. K. Etherington, *J. Mater. Chem. C Mater.*, 2022, **10**, 9484–9491.

34 H.-J. Zhang, R.-Q. Fan, X.-M. Wang, P. Wang, Y.-L. Wang and Y.-L. Yang, *Dalton Transactions*, 2015, **44**, 2871–2879.

35 M. Du, Y. Shi, Q. Zhou, Z. Yin, L. Chen, Y. Shu, G. Sun, G. Zhang, Q. Peng and D. Zhang, *Advanced Science*, DOI:10.1002/advs.202104539.

36 E. Regulska, R. Świsłocka, M. Samsonowicz and W. Lewandowski, *J. Mol. Struct.*, 2008, **887**, 209–215.

37 K. Pei, Y. Ma, X. Zheng and H. Li, *Chem. Phys. Lett.*, 2008, **457**, 323–328.

38 D. Bertoldo Menezes, A. Reyer, A. Benisek, E. Dachs, C. Pruner and M. Musso, *Physical Chemistry Chemical Physics*, 2021, **23**, 1649–1665.

39 Y. Tian, Y. Z. Shi, X. C. Fan, S. G. Wan, X. M. Wang, K. Wang, J. Yu, X. H. Zhang and C. Q. Ye, *Adv. Opt. Mater.*, 2023, **11**, 2300504.

40 J. V. Rival, S. Chand, A. Jana, Nonappa, V. Biju, T. Pradeep, P. Rajamalli and E. S. Shibu, *Advanced Materials*, 2025, **37**, 2507893.

41 I. E. Serdiuk, M. Mońka, A. Sikorski, K. J. Drozdowski, M. S. Eid, K. Wiśniewski, D. Trzybiński, M. E. Witkowski and W. Drozdowski, DOI:10.26434/CHEMRXIV.15001649/V1.

42 M. S. Najare, M. K. Patil, A. A. Nadaf, S. Mantur, M. Garbhagudi, S. Gaonkar, S. R. Inamdar and I. A. M. Khazi, *J. Mol. Struct.*, 2020, **1199**, 127032.

43 D. W. Chang, S.-J. Ko, J. Y. Kim, L. Dai and J.-B. Baek, *Synth. Met.*, 2012, **162**, 1169–1176.

44 G. F. Dvorko, E. A. Ponomareva, M. E. Ponomarev and M. V. Stambirsky, *Progress in Reaction Kinetics and Mechanism*, 2007, **32**, 73–118.

45 K. Anandhan, M. Cerón, V. Perumal, P. Ceballos, P. Gordillo-Guerra, E. Pérez-Gutiérrez, A. E. Castillo, S. Thamotharan and M. J. Percino, *RSC Adv.*, 2019, **9**, 12085–12096.

46 J. Casadot, T. M. Pappenfus, L. L. Miller, K. R. Mann, E. Ort, P. M. Viruela, R. Pou-Amérigo, V. Hernández and J. T. López Navarrete, *J. Am. Chem. Soc.*, 2003, **125**, 2524–2534.

47 S. C. Cevher and K. P. Pernstich, *Org. Electron.*, 2025, **136**, 107152.

48 G. P. Kissling, B. Ruhstaller and K. P. Pernstich, *Org. Electron.*, 2023, **122**, 106888.

49 S. Som, S. Dey, P. Ghosh, A. Ghosh, S. Mondal and D. Chopra, *Communications Chemistry 2025 8:1*, 2025, **8**, 330-.

50 T. Maadhu, RO. MU. Jauhar, A. Raja, S. Vediyappan, V. Gandhiraj, M. S. Pandian and P. Periyasamy, *J. Mol. Struct.*, 2023, **1294**, 136522.